\UseRawInputEncoding
\documentclass[%
 aip,
 amsmath,amssymb,
 reprint,%
]{revtex4-1}
\usepackage{graphicx}% Include figure files
\usepackage{dcolumn}% Align table columns on decimal point
\usepackage{bm}% bold math
\usepackage[T1]{fontenc}
\usepackage{mathptmx}
\usepackage{etoolbox}
\makeatletter
\def\@email#1#2{%
 \endgroup
 \patchcmd{\titleblock@produce}
  {\frontmatter@RRAPformat}
  {\frontmatter@RRAPformat{\produce@RRAP{*#1\href{mailto:#2}{#2}}}\frontmatter@RRAPformat}
  {}{}
}%
\makeatother
\begin{document}

\preprint{AIP/123-QED}

\title{A large octupole magnetic trap for research with atomic hydrogen}% Force line breaks with \\
%\thanks{Footnote to title of article.}%

\author{J. Ahokas}
\author{A. Semakin}
\author{J. J\"{a}rvinen}
\author{O. Hanski}
\author{A. Laptiyenko}
\author{V. Dvornichenko}
\author{K. Salonen}
\affiliation{Wihuri Physical Laboratory, Department of Physics and Astronomy, University of Turku, 20014 Turku, Finland}
\author{Z. Burkley}
\author{P. Crivelli}
\affiliation{ETH, Zurich, Institute for Particle Physics and Astrophysics, 8093 Zurich, Switzerland}
\author{A. Golovizin}
\affiliation{Lebedev Institute, 53 Leninsky pr., Moscow, Russia, Ru-119333}
\author{V. Nesvizhevsky}
\affiliation{Institut Max von Laue - Paul Langevin, 71 avenue des Martyrs, Grenoble, France, F-38042}
\author{F. Nez}
\author{P. Yzombard}
\affiliation{Laboratoire Kastler Brossel, Sorbonne Universit\'e, CNRS, ENS-PSL Universit\'e, Coll\`ege de France, 75252, Paris, France}
\author{E. Widmann}
\affiliation{Stefan Meyer Institute for Subatomic Physics, Austrian Academy of Sciences, Kegelgasse 27, A-1030 Wien, Austria}
\author{S. Vasiliev}
\email{servas@utu.fi.}
\affiliation{Wihuri Physical Laboratory, Department of Physics and Astronomy, University of Turku, 20014 Turku, Finland}

\date{\today}% It is always \today, today,

\begin{abstract}
We describe the design and performance of a large magnetic trap for storing and cooling of atomic hydrogen (H). The trap operates in the vacuum space of a dilution refrigerator at a temperature of 1.5 K. Aiming at a large volume of the trap we implemented the octupole configuration of linear currents (Ioffe bars) for the radial confinement, combined with two axial pinch coils and a 3 T solenoid for the cryogenic H dissociator. The octupole magnet consists of eight race-track segments which are compressed towards each other with magnetic forces. This provides a mechanically stable and robust construction with a possibility of replacement or repair of each segment. A maximum trap depth of 0.54 K (0.8 T) was reached, corresponding to an effective volume of 0.5 liters for hydrogen gas at 50 mK. This is an order of magnitude larger than ever used for trapping atoms.

\end{abstract}

%\pacs{67.80.F-, 67.80.fh, 76.30.-v, 07.07.Df - \textit{to be edited}}% PACS, the Physics and Astronomy
%Use showkeys class option if keyword
\maketitle

%\begin{quotation}
%Your lead paragraph
%\end{quotation}

\section{Introduction}
The invention of magnetic traps provided a broad range of possibilities for studies of neutrons and atoms at ultra-low energies without physically contacting material walls. These techniques combined with evaporative or laser cooling allowed reaching quantum degeneracy and boosted the research of cold atoms, including  precision spectroscopy and metrology. Several configurations of magnetic traps have been proposed in the middle of last century for trapping atoms \cite{Friedberg1951, Heer1963} and neutrons \cite{Vladimirski1960}. Pritchard \cite{Pritchard1983} emphasized the possibility of using 3D magnetic trapping for precision spectroscopy of neutral particles. He proposed to use an electric current configuration introduced by Ioffe\cite{Ioffe1961} for the stabilization of plasma: a set of linear conductors parallel to the trap axis (Ioffe-bars) combined with two pinch solenoids for closing the magnetic bottle in the axial direction. Since then the traps of this type are called Ioffe-Pritchard traps (IPT) and are widely used in experiments. The proposal of Hess\cite{Hess1986} to cool atomic hydrogen  by evaporating the most energetic atoms from an IPT promoted the quest for Bose Einstein Condensation (BEC) which was finally reached in 1995 for $^{87}$Rb\cite{Cornell1995,Ketterle1995} and three years later for atomic hydrogen\cite{Fried1998}.

IPTs developed for research with atomic hydrogen \cite{Doyle1991} and neutrons \cite{Yang2008} aimed at  large trap depths and volumes in order to maximize the number of trapped particles. More recently, progress in the production of slow anti-protons at CERN stimulated the development of large-size traps for antihydrogen, which were developed by ALPHA\cite{AlphaTrap2006} and ATRAP\cite{ATRAP2020} collaborations. The requirements of a large trapping field and volume are even more important there because of a very limited amount of antihydrogen available.

In this article, we describe the construction and test results of a large and relatively simple IPT designed for studies of ultra-cold atomic hydrogen by the GRASIAN collaboration\cite{GrasianPage}, the main goal of which is to observe and study gravitational quantum states (GQS) of H above the surface of superfluid helium\cite{LEAP2019}. We project formation of BEC in the potential well provided by gravity and the quantum reflection from the superfluid helium surface, as well as precision spectroscopy of the ultra-cold H gas. A large volume of the trap is important for the storage of a low density gas of H atoms at temperatures below 1 mK because it will increase the life-time of the sample limited by second order relaxation due to interatomic collisions. Another specific feature of experiments with atomic hydrogen is the necessity of a superfluid helium film lining the walls of the confinement cell, cooled to $\sim$100 mK by a dilution refrigerator. Since dry dilution refrigerators with a common vacuum space are becoming standard  equipment for desktop experiments, we designed our IPT for operation in such conditions.  We found nearly equal performance of the trap in the bath of liquid helium at 4.2 K and in vacuum being cooled by the 1 K pot of the dilution refrigerator. We reached the maximum trap depth of 0.8 T, similar to that in large scale installations\cite{AlphaTrap2006, ATRAP2020}, while having nearly the largest effective volume: 0.5 liters at the gas temperature of $T=50$ mK. Our trap has a modular construction with an octupole magnet for  radial confinement, two axial pinch coils and a 3 T solenoid for the hydrogen dissociator. All coils, including the 8 race-track modules of the octupole magnet, are assembled so that each coil may be easily replaced in case of a possible damage, e.g. caused by a catastrophic quench. 

We discuss below the choice of the magnetic field configuration, trap design, details of construction, manufacturing procedure, and test results in a liquid helium bath as well as in the vacuum of the dilution refrigerator. We also discuss possible improvements of the trap design for reaching stronger trapping fields and faster field ramping.

\section{Design considerations}
\begin{figure}[t]
    \includegraphics[width=8.5 cm]{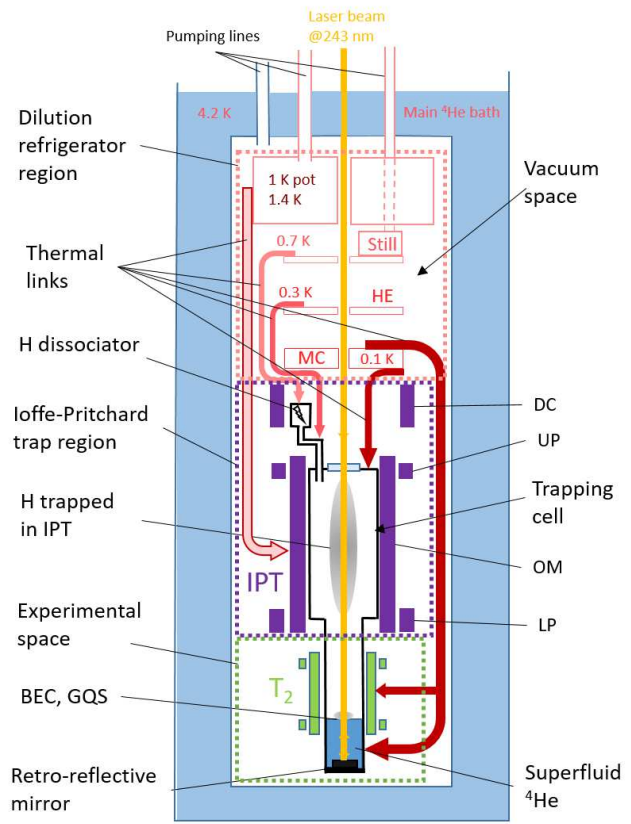}
    \caption{Schematics of the experimental setup for experiments with ultra-cold atomic hydrogen. DC - dissociator solenoid; UP - upper pinch; OM - octupole magnet; LP - lower pinch. Main components of the dilution refrigerator are shown: 1 K pot; Still; heat exchangers (HE); mixing chamber (MC). A laser beam at 243 nm for two-photon 1S-2S spectroscopy passes along the setup axis and is retro-reflected by a mirror at the bottom. Thermal links between different components of the dilution refrigerator, the traps and other experimental components are shown with red arrows.}
    \label{fig:ExpSetup}
\end{figure}

Research with hydrogen atoms at the lowest energies projected by the GRASIAN collaboration\cite{GrasianPage, LEAP2019} requires a large number of hydrogen atoms at temperatures of $\leq 1$ mK, and sufficiently long storage times on the order of tens of minutes. The low energy atoms will be released from the trap or transferred to another, more shallow trap ($T_2$ in Fig.\ref{fig:ExpSetup}) for further manipulations in the phase space. This defines our major considerations related to the choice of the trap parameters. The main reason for the loss of  trapped hydrogen is the process of  dipolar relaxation which occurs in binary collisions of the atoms and causes a spin flip into the untrapped state\cite{Lagendijk1986}. This is a process of  second order in density of atoms, and its rate can be minimized by decreasing the gas density. The characteristic decay time of the H sample due to this process is $\tau_{rel}=(G_{rel}\times n)^{-1}$. For a rate constant $G_{rel}\approx 10^{-15} $cm$^{3}$/sec\citep{Lagendijk1986} and a gas density $n=2\times10^{12} $cm$^{-3}$ we obtain $\tau_{rel}\approx 500$ sec. To get a large number of atoms at small densities, the trap volume has to be large. Since the existing sources of H atoms based on cryogenic dissociators may provide atomic fluxes exceeding $10^{13}$ atoms/sec, loading the trap with up to $\ge 10^{15}$ atoms can be done in a reasonably short time, $\approx100$ sec. Trapping fields of about $0.7$ T are sufficient for a tight enough confinement of the H gas at $T=100-200$ mK. We estimate that a subsequent forced evaporative cooling will result in about $10^{14}$ hydrogen atoms at a temperature of $\leq 1$ mK. These values present a significant improvement over previous works \citep{Walraven1996,Fried1998}, and were taken as goals for the design of our IPT.

Ioffe-Pritchard traps used for studies of cold hydrogen by the MIT \cite{Doyle1991,Fried1998} and Amsterdam \cite{VanRoijen1989,Walraven1996} groups had quadrupole configurations with four Ioffe bars for the radial confinement. Quadrupoles feature a linear dependence of magnetic field in the radial direction with the best concentration of atoms at the trap bottom. This raises the critical temperature of the BEC \cite{Kleppner1987} at the expense of the decreased effective volume of the trap compared to the configurations with larger numbers of Ioffe bars $N_p$. In general, IPTs provide a radial confinement potential of the form $U(r)\sim B(r)\sim r^{N_p/2-1}$. Increasing $N_p$ leads to a more homogeneous field in the central region of the trap and a steeper growth near the wall. Thus the octupole configuration with $N_p=8$ bars provides a dependence of $B(r)\sim r^3$. The effective volume of the octupole trap is $\sim 5$ times larger than that of the quadrupole for the same ratio of the trap depth to the gas temperature. Due to this reason, the octupole geometry has been chosen for trapping antihydrogen \cite{AlphaTrap2006,ATRAP2020}.

Strong superconductive magnets are widely used in accelerator and nuclear physics for focusing neutral particles. The coils of such focusing systems are typically wound with multi-wire superconductive (SC) cables. Several tens of ordinary multi-filament SC wires are weaved together in such a cable. This arrangement simplifies winding, but requires high energizing currents. Quadrupole magnets developed by KEK (National Laboratory for High Energy Physics, Japan) for focusing electrons in the TRISTAN project were designed for currents of $\sim 4$ kA \cite{KekIEEE1991}. Another advantage of high current magnets is a smaller inductance which allows faster ramping of magnetic fields. This was very important in experiments with ultra-cold neutrons \cite{KekCryogenics2014}. The high-current approach was used in the octupole magnet of the ALPHA collaboration for trapping antihydrogen \cite{AlphaTrap2006} where the nominal operating current was 1 kA. High currents require special cryogenic solutions for the current leads and strongly complicate the cryostat construction \cite{KekCryogenics2014}. This approach is difficult to realize for small labs and desk-top experiments. Our choice is low current magnets wound with standard SC wires available from multiple suppliers. 

Wall-free confinement of atoms in magnetic traps is the main feature which allows to stabilize and cool atoms. Invention of laser cooling provided an easy way of loading magnetic traps for most alkalis, running experiments in a room temperature environment. However, experiments with hydrogen are different. Laser cooling is very difficult and inefficient due to the VUV wavelengths required for the excitation from the electronic ground state. There is no other solution to pre-cool the H gas to trappable energies  than to use cryogenics and dilution refrigerators (DR). The flux of low-field seeking H atoms (H$\uparrow$) from a cryogenic H$_2$ dissociator operating typically at $\sim0.7$ K is cooled by thermalization with the physical walls. The dissociator is located at the fringing field of a strong solenoid $B\sim 3.5$ T where a negative field gradient pushes H$\uparrow$ towards the trapping cell. In order to avoid adsorption and subsequent recombination, the walls of the line between the dissociator and the trapping cell need to be covered with a superfluid helium film. A dilution refrigerator is used for cooling different parts of this line, finally cooling the gas to $100-200$ mK. The gas in the trapping cell and the superfluid helium film should be isolated from the vacuum space of the DR. Construction and materials of the trapping cell should provide high enough thermal conductivity while minimizing heating due to eddy currents during magnetic field ramping. 

Hydrogen gas pre-cooled in the large IPT described in this paper will be used for several experiments including spectroscopy of GQS above the surface of superfluid helium\cite{LEAP2019}. Formation of a free surface of superfluid helium is only possible in the lowermost region of the whole setup below the IPT (see Fig. \ref{fig:ExpSetup}). Cooling this region to $\leq100$ mK requires bringing a thermal link from the DR to this region. Satisfying all the above mentioned requirements presents a serious challenge. A reasonable solution was found by placing the IPT, trapping cell, and GQS experiment components inside the vacuum can of the DR. The IPT magnets are cooled by the 1K pot of the dilution refrigerator (Fig.\ref{fig:ExpSetup}). 

  \subsection*{Specifications summary}
In summary, our Ioffe-Pritchard trap is designed to meet the following specifications:
\begin{itemize}
    \setlength{\itemsep}{0pt}
    \item Minimum trap depth 0.8 T ($\approx$0.54 K);
    \item Effective volume for H gas at 50 mK $\geq$ 500 cm$^3$; 
    \item 3 T dissociator solenoid located above the upper pinch coil;
    \item Operation at 1.5 K in the vacuum can of the dilution refrigerator;
    \item Length of the system $\leq 0.6$ m, fitting in the diameter $\leq210$ mm;
    \item Field ramping  to 90$\%$ in $\leq 500$ sec;
    \item Operating currents $\leq$100-120 A;
\end{itemize}
\section{Construction of the trap}
\subsection{Octupole magnet (OM)}
Each Ioffe bar which provides magnetic field for radial confinement should have a return path for closing the current loop. In most IPTs the trap length along the axis is much larger than its diameter. Therefore, the radial confinement is typically provided by elongated coils in the form of a race-track (RT). Currents in the neighbouring bars should be in opposite direction. There are several possible ways to arrange the RT coils to obtain such a configuration. 
\begin{figure}[t]
    \includegraphics[width=\columnwidth]{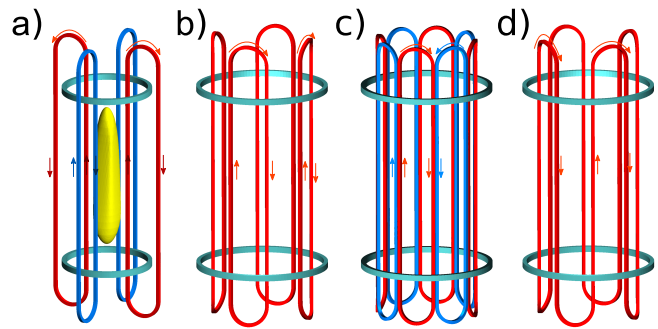}
    \caption{Configurations of currents for Ioffe-Pritchard traps of different types: a) race-track (RT) coils located in planes going through the main IPT axis \cite{VanRoijen1989}; b) RT coils located symmetrically around the axis, each Ioffe bar is a single RT linear current; c) the same as b) but with the double number of RTs. Each Ioffe bar is formed by two linear currents of neighbouring RTs \citep{Doyle1991,Yang2008}; d) the construction is not based on RT coils. Ioffe bars are formed by the linear parts of the serpentine-like currents wound on a cylindrical surface (mandrel) \citep{AlphaTrap2006,ATRAP2020}}
    \label{IPT configurations}
\end{figure}

\textit{a)} RT coils are arranged  diametrically in planes containing the trap axis, so that the return path of each Ioffe bar is taken far away from the trap axis and has a negligible influence on the trapping field (Fig. \ref{IPT configurations} a).  Such configuration was used for trapping H in Amsterdam \cite{VanRoijen1989}. This design increases the radial size of the system and complicates the arrangement and winding of the pinch coils. It would be difficult to build a trap with a large inner bore diameter using this scheme.

\textit{b)} Race-tracks are arranged with the current bars at the equal distance from the trap axis, e.g. all the RT coils are located on the same cylindrical surface (Fig. \ref{IPT configurations} b). Since each RT coil has two linear current bars, $N_p/2$ coils are sufficient to build a $N_p$-poles magnet system for the radial confinement. Thus four RT coils may be arranged symmetrically around the system axis to form the octupole trap. In this case, as can be seen in Fig. \ref{IPT configurations} b, the turn-around pieces of the race-tracks produce non-zero fields on both ends of the system. These fields have opposite directions in the upper and lower ends of the octupole magnet, implying that there is a zero field region in the trap center where the Majorana relaxation to  untrapped states may lead to the rapid loss of the sample. The problem can be solved by applying a bias field to the whole trap. The final field configuration becomes asymmetrical in the axial direction, and the effective volume of the trap is reduced.

\textit{c)} A similar construction as in \textit{b)}, but using twice as many RT coils, allows obtaining a fully symmetrical field configuration. Now each of the Ioffe bars consists of two linear RTs with parallel currents, and 8 RTs are required to build the octupole configuration (Fig. \ref{IPT configurations} c). The end-pieces of the RTs have alternating current directions, and do not produce any field on the trap axis. Quadrupole traps built for trapping atomic hydrogen \cite{Doyle1991, Fried1998} and neutrons \cite{Yang2008} utilized such a scheme with four RT coils. This configuration allows better mechanical stability against the large magnetic forces between the linear pieces of the RT coils.

\textit{d)} Ioffe bars can be produced by winding SC wires or cables on a cylindrical surface in a serpentine way (Fig. \ref{IPT configurations} d). This arrangement requires a special technique with robust fixing of each winding on a special form. Using a small diameter wire with a large number of turns makes the winding procedure extremely complicated and expensive. Winding with bundles of wires makes the process somewhat easier, but leads to an increase in the trap operating current. Such trap constructions were used to build octupole traps for the ALPHA \cite{AlphaTrap2006} and ATRAP \cite{ATRAP2020} collaborations for antihydrogen at CERN. The octupole segments in these traps cannot be disconnected from each other and if something (e.g. catastrophic quench) happens to one of them, the whole radial magnet system should be changed or re-wound. 

\begin{figure}[t]
    \includegraphics[width=8.5 cm]{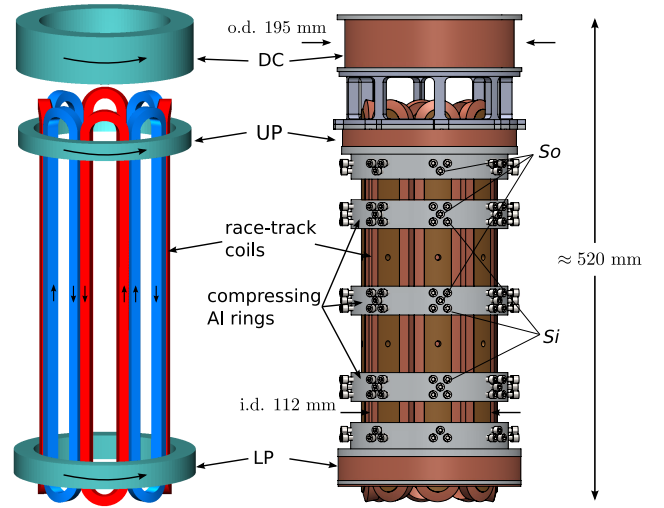}
    \caption{The configuration of currents and coils chosen for the Turku IPT trap (left); an IPT outlook with all coils and compressing Al rings (right). The octupole magnet inner bore is 112 mm, its outer diameter is 150 mm. The overall height of the IPT including the dissociator solenoid is $\approx$ 600 mm. The race-track coils of the octupole magnet are put together with three 1 cm thick Al compressing rings in the center of the system plus two rings extending from the pinch coil forms. Tightening the M8 bolts "$So$", threaded through the compression rings into each race-track mandrel, moves the race-tracks out of the axis. The M5 screws "$Si$", four for each compression ring and two for each pinch ring, press the race-tracks inwards. They were tightened with 2 N$\cdot$m torque, preventing possible motion of the race-tracks due to magnetic forces directed outwards.}
    \label{IPT outlook}
\end{figure}

To meet the required specifications for our IPT, we selected the type c) octupole configuration with eight RT coils which, together with the cores, form a thick-walled cylinder. The RT segments are assembled and pressed together in a way similar to the staves of an old wooden barrel, as shown in Fig. \ref{IPT outlook}.
    
\subsection{Axial solenoids}
Two solenoidal coils are required to provide axial confinement in the IPT trap. They are denoted as upper (UP) and lower (LP) pinch coils. The pinch coils provide two field maxima located close to the turn-around regions of the octupole coils. For atomic hydrogen one more coil (DC) with a 3-4 T field is needed to accommodate a cryogenic dissociator \cite{Doyle1991, VanRoijen1989}.

Construction and manufacturing of these simple solenoidal coils is easy, and the pinches could generate much larger fields than the specified trap height. However, the fields of the pinches and of the octupole have different symmetries. Their vector sum varies considerably in the trapping volume. The optimization of the trap depth becomes a complicated task requiring calculation of the total field in the volume of the trap. We performed such calculations by numerical integration\cite{BiotSavart} based on the Biot-Savart law. Dimensions of the pinch and dissociator coils and their location with respect to the octupole center were varied to find the optimum configuration. The mutual arrangement of the IPT coils with the current directions is schematically shown in Fig. \ref{IPT outlook}.  

\begin{figure}[t]
    \includegraphics[width=\columnwidth]{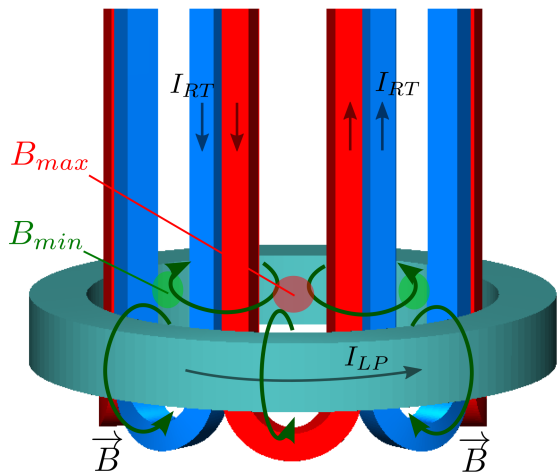}   
    \caption{The configuration of currents and magnetic fields near the edge of the lower pinch in the so-called "leakage spots" of the trap. Due to the alternating direction of the field generated by the octupole segments, for every second race-track, the radial component of this field is subtracted from that of the pinch. These regions near the upper(lower) edge of the lower (upper) pinch define the absolute minimum of the trapping field $B^{wall}_{min}$ on the wall of the trapping cell.}
    \label{Leakage regions}
\end{figure}

The pinch coils cannot be very long due to the overall length limitation, while their diameter should be large enough to accommodate the whole octupole system. Typically for IPT traps, the diameter of the pinch coils is larger than their length. This means that the field near the edge of such a short coil is larger than the field in the center. The field near the edge has a large radial component which interacts constructively or destructively with the octupole field (Fig.\ref{Leakage regions}) since the latter changes direction for every second octupole segment. The areas in the cylindrical surface of the bore of the octupole, where the radial field of the pinch is subtracted from the octupole field, are the possible trap leakage spots where the trap barrier minimum $B^{wall}_{min}$ is located. The IPT trap should be designed to have sufficiently strong field in these regions. 

The dissociator solenoid geometry was chosen to obtain the field of $\ge$3 T near its inner wall.  The fringing field of the dissociator solenoid adds to the field of the upper pinch. This allows a substantial reduction of the size of the upper pinch and its operating current. 

\subsection{Magnetic forces}
Considerations of magnetic forces in the IPT construction are extremely important. The force experienced by the linear part of the race-track in the field of 0.5 T perpendicular to the bar current of $8.5\times10^4$A (250 A$\times$340 turns) is $\approx$ 15 kN. Such repulsive forces between the linear parts of individual race-tracks were reached during their tests and training at maximum currents. However, in the octupole configuration with 8 race-tracks shown in Fig. \ref{IPT outlook}, each Ioffe bar is formed by two linear conductors of the race-tracks carrying parallel currents. Attractive forces between these current bars (red arrows in Fig. \ref{RT CrosSec}) have no radial component and create a large compressive stress which  binds them strongly together. The net force experienced by such Ioffe bar from the other 7 bars in the central region of the trap is directed outwards (blue arrow in Fig. \ref{RT CrosSec}). For the desired operating conditions with the trap depth of 1 T we obtain $\approx460$ N per 10 cm of the bar length. This corresponds to the pressure of $\approx0.14$ bar and is much smaller than the tangential compressing forces between the linear RT parts. This means that for our construction of the octupole, the magnetic forces make the whole assembly mechanically stronger and the octupole can operate without any extra stressing force from outside, except for a weak binding to keep the segments together at zero current. This pre-stressing force is provided with the $Si$ bolts screwed into the compressing rings (Fig. \ref{RT CrosSec}). Enhancement of the mechanical strength of the octupole due to its own magnetic forces was confirmed in the tests of the IPT. 

Interaction between the turn-around sections of the race-tracks and the pinches creates another problem. Currents in these end-pieces change direction from parallel to anti-parallel with respect to the pinch solenoid. We performed a numerical calculation of this force  using Biot-Savart software \cite{BiotSavart}. For our IPT, the net force to each curved region is fairly large $\approx 2$ kN. The stability of these turn-around regions is also improved by the strong attractive forces between the linear RT parts. In addition, we used the bolts $Si$ screwed to the pinch cores and three compressing Al rings (Figs. \ref{IPT outlook}, \ref{RT CrosSec}). The  bolts press onto the bronze mandrels upon tightening and provide an extra force on the RTs directed towards the trap axis. 

\section{Manufacturing procedure}
Each of the 8 race-track coils of the octupole, two axial pinch solenoids and dissociator solenoid were wound separately on its own mandrel. For the race-track mandrels we used bronze and aluminium alloy for the axial coils. Thermal expansion of these materials is not much different from the composite winding of the coils and both were readily available from   local suppliers. Race-track mandrels were milled from 142 mm o.d. bronze tube with a wall thickness of 12 mm. Side surfaces adjacent to the linear parts of the race tracks were milled with such an angle that their planes intersect approximately on the tube axis (see Fig. \ref{RT CrosSec}).
 \subsection{Coil winding}
All coils were wound with  0.55 mm thick (bare dia. 0.50 mm) Luvata OK-54 superconductive wire\cite{Luvata} which contains 54 NbTi filaments of 45 um diameter in copper matrix with the nominal Cu to SC ratio 1.33 and outer diameter of 0.5/0.55 mm bare/insulated. The short sample critical current is 206/340 A in a field of 5.4/3.0 T at a temperature of 4.2 K.
 
\begin{figure}[t]
    \includegraphics[width=\columnwidth]{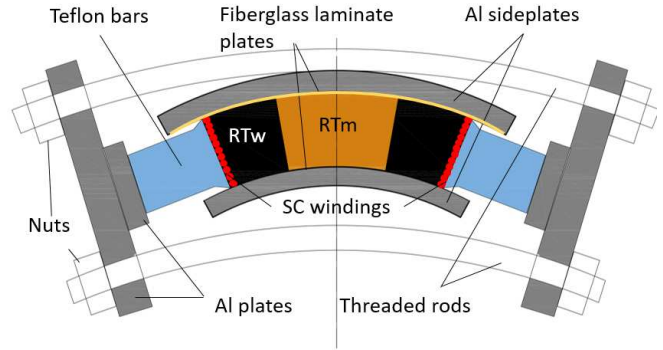}
    \caption{Pressing tool for linear parts of the race-track coils. RTw - race-track windings; RTm - race-track mandrel; SC windings - superconductive wire windings}
    \label{RT pressing}
\end{figure}

For winding the race-tracks, 5 mm thick aluminium side plates were bolted to the bronze mandrels from the inner and outer curved sides (Fig. \ref{RT pressing}). The inner/outer surfaces of the plates have the same curvature as the outer/inner surfaces of the mandrels and the desired curvature of the octupole system. Fibreglass laminate plates of 0.3 mm thickness were placed between the aluminium side plates and the mandrels.

All coils were wet wound using Stycast 2850FT epoxy with Catalyst 11. The bobbin with the SC wire was loaded with a weight keeping $\sim80$ N tension along the wire during winding. However, this tension is not transferred to the linear parts of the race tracks. The wires there remain nearly free and require additional compression. In order to provide sufficient strength to the linear parts and achieve homogeneous filling by epoxy, we pressed the linear parts from both sides using teflon and aluminium bars screwed together with 8 M6 bolts providing a total force of $\sim$ 1 kN onto the winding layers (Fig. \ref{RT pressing}). For the reinforcement and extra insulation of the coil windings we used Zylon HM273 filament\cite{Zylon}. The winding procedure of the RT layer was as follows: a turn-to-turn winding of a full layer (19 turns), painting by epoxy with a brush, winding a Zylon filament layer (between the SC wires) on top of the epoxy, and application of extra epoxy to cover evenly all surface of the layer. After winding 2-4 layers, the pressing bars were installed and screwed together as shown in Fig. \ref{RT pressing}. Excess epoxy was removed from the turn-around regions, and the race-track was moved to the oven for curing at 85 $^{\circ}$C. Usually, 2-4 layers were produced per day, followed by an overnight curing. Each race-track consists of 18 layers. In total we manufactured 10 nearly identical race-tracks, eight of which were assembled together in the octupole magnet, and two kept in stock for a possible replacement/repair.

The pinch and dissociator solenoids were wound using the same tension, epoxy and Zylon reinforcement. Since no extra compression was required, 4-5 layers could be done during a day followed by overnight curing. The winding numbers of all coils are presented in Table \ref{tab:coils}.
\begin{table}[h!]
  \begin{center}
    \caption{IPT coil parameters: $N_w$ - number of windings; $L$ - inductance, H; $I_c$@4.2 - critical current reached in individual coil tests at 4.2 K, A; $I_c$@1.4 - critical current reached in individual coil tests at 1.4 K, A; $I_t$ - calculated current for reaching the target trap depth of 0.8 T for the simultaneous operation of all coils, A}
    \label{tab:coils}
    \begin{tabular}{m{4em}|m{4em}|m{4em}|m{4em}|m{4em}|m{4em}}
     \textbf{Coil} & \textbf{$N_w$} & \textbf{$L$} & \textbf{$I_c$ }@4.2 & \textbf{$I_c$}@1.4 & \textbf{$I_t$}\\
      \hline
      RT & 340 & 0.04 & 240$\pm10$ & &\\
      OM & 2720 & 0.43 & 173 & 166 & 120\\
      LP & 1400 & 0.53 & 120 & 131 & 92\\
      UP & 760 & 0.2 & 170 & $>100$ & 32\\
      DC & 3450 & 2.4 & 130 & 130 & 100\\
    \end{tabular}
  \end{center}
\end{table}

\subsection{IPT assembly}
\begin{figure}[t]
    \includegraphics[width=\columnwidth]{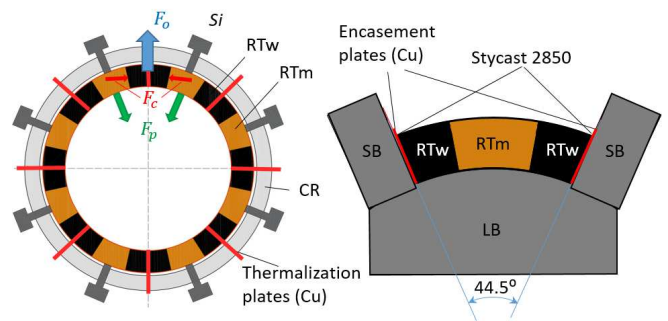}
    \caption{Cross section of the octupole magnet assembly with the race-track coils and Al ring (left). Cross section of the final race-track shaping tool (right). \textit{Si} - pressing bolts, CR - Al compression ring; SB - Al side-bar. LB - Al lower bar. Magnetic forces acting of the RT bars are shown on the left picture with the arrows: red arrows $F_c$ - for the attractive force between the flat RT regions; green arrows $F_p$ - mandrel pressing forces from the \textit{Si} screws; blue arrow $F_o$ directed outside - the net force acting from the other 7 current bars.}
    \label{RT CrosSec}
\end{figure}

The winding procedure described above does not ensure accurate enough final dimensions for the race-track coils. In order to get a good, tight and  symmetrical assembly of the octupole segments, we made a final processing of their linear parts for each RT matching the angular segment of $\approx44.5^{\circ}$. This angle is slightly smaller than 45$^{\circ}$ in order to leave space for extra thermalization plates between the RTs. The RTs after winding, together with 0.3 mm thick encasement copper plates, were put into a special form (Fig. \ref{RT CrosSec}). Stycast 2850 epoxy was applied to fill the gap between the linear RT parts and the encasement plates, and cured. Putting together the RTs, we installed extra 0.5 mm thick thermalization plates pressed between the encasement plates. The octupole assembly forms a tubular construction with an outer diameter of 137 mm and and a clear bore of 111 mm. Pinch coils and dissociator solenoid cores are fixed on the OM using the $So$ bolts (Fig. \ref{IPT outlook}) screwed to the RT mandrels via the holes in the Al cores of the pinches and three compressing rings located between the pinches. The $Si$ bolts (16 per each RT) screwed in the Al compressing rings press onto the RT cores and provide a pre-stressing force on them directed towards the IPT axis. Final adjustment of the tubular shape was done by loosening the $So$ and tightening the $Si$ bolts using for the latter a final torque of $\approx2$ N$\cdot$m .

\begin{figure}[t]
    \includegraphics[width=\columnwidth]{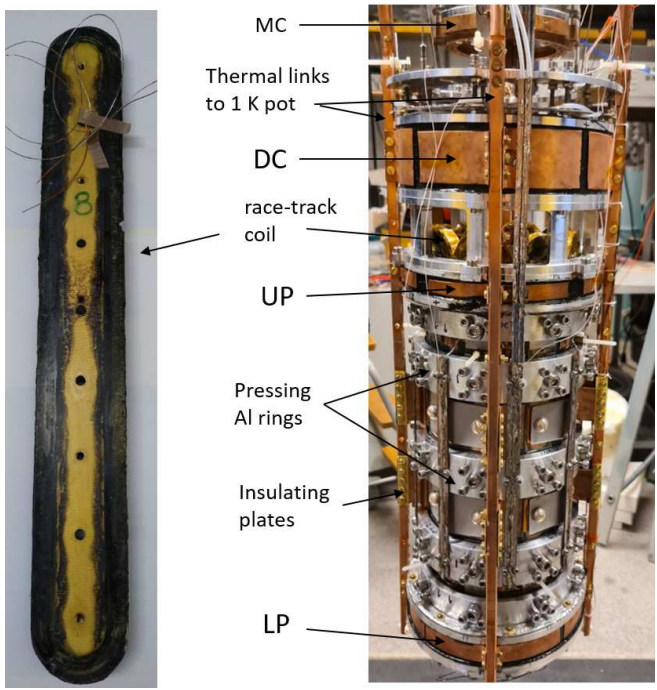}
    \caption{Photographs of an individual race-track coil just after winding and before finalizing its shape (left) and of the final IPT assembly hanging in the dilution cryostat (right).}
    \label{IPT photos}
\end{figure}

The RT thermalization plates and the RT mandrels were thermally linked to four 12 mm diameter copper rods going up and are attached to the 1 K pot of the cryostat (1 K bars). All axial solenoids were thermalized using altogether 8 curved copper plate segments glued to the coil outer surface (Fig. \ref{IPT photos}) and linked to the 1 K bars (Fig. \ref{IPT photos}). The total length of the 1 K bars from the IPT bottom to the 1 K pot is $\approx$1.4 m. Special attention was paid to avoid closed current loops through the thermal links. Eddy currents induced in such loops could cause excessive heating during the field ramps. All connections between the copper plates and vertical bars were done via thin insulating laminate plates or 30$\mu$m thick Kapton films.

\subsection{Current leads}
Wire connections of the race-track coils were done by soldering the wires to a copper plate over the length of $\approx 30$ cm. We were aware that even for such a long soldering, joint resistances may cause a substantial overheating of the magnet system under operating conditions in vacuum. In order to avoid heating of the joints, we made electrical connections utilizing the cold welding techniques of the SC strands of the wires\cite{SCpressing}. The SC filaments were freed from the copper matrix by etching in nitric acid. The strands from two ends were twisted together and pulled into a small piece of copper tube with tight packing. Finally, the end was pressed with a 20 kN force. Current leads of all the IPT components from the coils up to the top of the vacuum can of the DR were built from the same SC wire as was used for the coils, using the same cold welding joint technique. 

The SC wires were fed through the vacuum feedthroughs on the top flange of the vacuum can and soldered to the leads going  through the helium bath of the cryostat up to room temperature. The lower end of the leads in the cryostat neck and helium bath were built from commercial High-Temperature Superconductor tapes \cite{HTStapes}. We built four pairs of leads, one for each of the IPT components: the octupole (nominal current of 150 A), the lower pinch (130 A), the upper pinch (100 A), and the dissociator coil (100 A).

We used passive quench protection which was done with protection resistors connected in parallel to each of the IPT coils made from 0.8 mm dia. pieces of stainless steel wire. Their resistance was $\sim1 \%$ of the coil resistance at room temperature. Then, during a quench $\sim90\%$ of the stored coil energy should be released in the protection resistor. The resistors were placed in the vapour of the helium bath of the DR cryostat, just above the maximum helium level. They were soldered to the lower ends of the HTS leads. This allows a substantial reduction in amount of liquid helium evaporated in a quench. 

\section{Coil training and performance tests}
Each RT was tested and trained in liquid helium at 4.2 K in a small cryostat. Typically, we had the first quench at 130-150 A, reaching 230-250 A after 5-8 quenches. For a single race-track coil, the maximum field is reached near the inner surface of the turn-around ending. With 250 A current the maximum field of $\approx 4$ T was reached. This is $\approx 85\%$ of the short sample critical current. The tests of the individual race-tracks were done without any extra support, training the coils just on their mandrels as they came out after manufacturing. At the maximum currents, the repulsive force between the linear RT parts reaches 4.3 kN per 10 cm of length. To our surprise, all coils withstood such force and no damage was observed. This proves a very good mechanical strength of the wire bars re-enforced with Zylon, and a proper choice of the winding procedure. Being immersed in liquid helium, the race-track coils allowed fairly high field ramping rates. Energizing to 150 A current could be done in 50 sec.

Measurement of the magnetic field was performed with an axial cryogenic Hall probe HGCA-3020 (Lake Shore Cryotronics\cite{HallProbe}) fixed at the distance of $\approx$ 5 mm from the center of the linear part of the race-tracks. The measurements provided $\approx10\%$ scatter for different RTs, matching the calculated value with the same accuracy. Such an error can be explained by inaccuracy in the probe positioning.

\subsection{Tests of the IPT in liquid helium at 4.2 K}
After testing the race tracks, the whole IPT was assembled and pre-stressed as described above. The first two performance tests of the whole system were done in the main helium bath of the DR cryostat at 4.2 K. The octupole assembly was tested and trained first. The maximum current of 173 A was reached after four quenches with the first one occurring at 130 A. Then, the axial coils were tested separately. The lower pinch reached 130 A, the upper pinch 170 A, and the dissociator solenoid 130 A after 2-3 quenches. These maximum values were reproducible within 10-15$\%$ for the different test runs.

Charging all coils together, we aim at reaching the maximum trap depth $\Delta B_t = B_{min}^{wall} - B_{min}$ defined as the difference between the field minimum $B_{min}^{wall}$ at the inner wall of the cylindrical trapping cell ($r=53$ mm) and the absolute field minimum $B_{min}$ in the trap center. $B_{min}^{wall}$ is located in the IPT leaking spots as described above. By a numerical calculation, we found the following current combination which allowed us to reach the desired trap depth $\Delta B_t = 0.8$ T: 120/ 92/ 32/ 100 A for the currents in the OM/ LP/ UP/ DC coils respectively, which we set as the target of the IPT performance. Desired trap depth in the upper end of the IPT can be reached with different combinations of the UP and DC currents since the fields of these coils add together below the UP. However, $\ge$100 A current in the DC coil is required to ensure $\ge$3 T field for the H dissociator as specified for our IPT. This leads to a fairly low current in the UP. Below, we will present results of the tests for the various coil combinations and report currents relative to the specified above target values $I_t$.

Testing the coils together, we found that if we charge the lower pinch first to 80-90 A and then try to energize the octupole, a quench occurs at currents well below 100 A. Changing the charging order so that first the octupole is loaded first to $\sim$100 A, allowed reaching currents above 100 A in the lower pinch. Increasing further the OM current, the steady operation of both coils was reached with 130/ 100 A currents in the OM/ LP combination (108$\%$ $I_t$). The observed dependence of the performance on the charging order confirms that the magnetic forces acting between the race-tracks improve the stability of the whole system.

Tests of the upper axial coils together with the octupole showed a slightly better performance. We reached currents of 138/ 40/ 115 A for the OM/ UP/ DC combination ($115\%$ $I_t$). 

Finally, all coils were energized together, raising first the octupole current to 100 A, then charging all others to $\sim 70\%$ of $I_t$ and further increase of all currents in proportion to the target values. Finally, we reached 130/ 100/ 40/ 105 A for the OM/ LP/ UP/ DC simultaneous operation (108$\%$ $I_t$). The currents reached in the 4.2 K tests are listed in Table \ref{tab:coils}.

The magnetic field was measured in the center of one of the race-track bronze cores at the distance of $\sim$ 52 mm from the axis using the mentioned above Hall probe. The results agreed with the calculations to within about $10\%$ accuracy.

Several tens of quenches were observed altogether during the "wet" tests in helium. Voltage taps were connected to each segment of the octupole and to each axial coil. Recording the voltages over the taps during quenches allowed identifying which coil quenches first and triggers transition of the whole system. In the liquid helium tests, sometimes one or two of the coils remained superconducting and energized while the others quenched. The OM and the LP recovered a few seconds after the quench with a total evaporation of 5-7 liters of helium at the maximum current. The DC quench was performed only twice. This coil has the largest stored energy and its quenches evaporated more helium, about 10 liters. 

\subsection{Tests of the IPT in vacuum at 1.5 K}
After the tests in liquid helium, the IPT including the 1 K bars and current leads was mounted in the vacuum space of the dilution cryostat (see Fig. \ref{IPT photos}). The whole assembly was centered in the 210 mm i.d. vacuum can of the dilution refrigerator. Several thermometers were fixed to every IPT coil and to the magnet leads in the helium bath. With zero currents all IPT coils reached  $\approx1.5$ K, very close to the temperature of the 1 K pot.

As expected, the main difference of the tests in vacuum was a substantial heating of the coils during the current ramps. This was caused by the eddy currents in the coil mandrels and in possible parasitic loops between the thermal links and 1K bars. In vacuum we had to reduce the ramping speed by nearly an order of magnitude compared with the tests in liquid helium. Charging any of the coils alone to $90\%$ of the target required $\sim$10-15 min and caused a heating of the IPT to $\approx 3.5$ K. Analysing the temperature gradients during the field ramps, we came to the conclusion that the bottleneck in the heat transfer is caused by poor thermal contact between the 1 K bars and the 1 K pot. Several simple solutions to improve this contact will be discussed below.

We expected that quenches in vacuum may cause more serious troubles than in liquid helium and a much higher risk of IPT damage. Therefore, for the tests in vacuum, we did not make extensive training of the whole IPT. Since the critical current of the SC wires increase by $\sim30 \%$ at 1.5 K, and no changes were done for the mechanical connections of the IPT components, we expected that no quench should occur before we reach the same critical currents as in the wet tests. However, since the system behaviour at quench is an important experience, individual IPT components and their simple combinations were quenched for a few times.

First we ramped the LP and then the OM to quenches. The former was quenched twice, and the latter four times. We have not observed training behavior, and the maximum currents were within 5$\%$, the same as in liquid helium at 4.2 K. Each quench boiled off all ($\approx 3$ l) helium in the 1 K pot and warmed the magnet system up to 30-40 K. In order to facilitate fast recovery, a small amount of exchange gas was admitted in the vacuum space establishing a good thermal contact to the main helium bath. The 1K pot and the IPT cooled back and were ready for the next ramp after $\approx$ 30 minutes. 

The tap voltages across each RT of the OM and the LP were monitored during quenches. No "bad" RT with reproducible first quench behaviour was detected, indicating that they are all of about the same quality, and so far none of them need a replacement. Next, we tested the UP and DC, energizing them to 100 A and 130 A respectively. Stable operation without quenches was observed. Since these currents well exceeded the target values, we did not make an attempt to go further.  

Then, the OM and the LP combination was tested. We charged the OM to 80$\%$ of the target current, then ramped the LP to the same rating, then slowly increased both currents. We reached $106\%$ of $I_t$ with continuous operation during 30 min and no quench was observed. 

Then, the OM, UP and DC were tested together, ramping first the OM to $\sim$100 A. We reached the currents 138, 40 and 115 A (115$\%$ $I_t$) respectively for these coils when the quench occurred. This quench had the strongest effect with largest overheating and recovery time since the largest part of the magnetic energy is stored in the DC coil. No damage occurred after this event. The system recovered to the initial state and we were able to continue the tests.

Finally, all the IPT coils were energized to 100$\%$ of the target current and kept for 15 minutes without a quench. We did not keep all currents in the IPT longer than that because continuous operation warmed the current leads approaching critical values for the HTS tapes. On the other hand, longer operation times are not required in experiments with atomic hydrogen. We expect that the trap will be loaded with hydrogen atoms for 5-15 minutes. Possible improvement for reaching longer continuous operating times are suggested below. We have not observed any noticeable heating of the IPT assembly at the steady current state when the field ramps were stopped. This means that the soldering and cold welding connections are good enough and do not impose heating problems.

\section{Magnetic field profiles}
\begin{figure}[t]
    \includegraphics[width=\columnwidth]{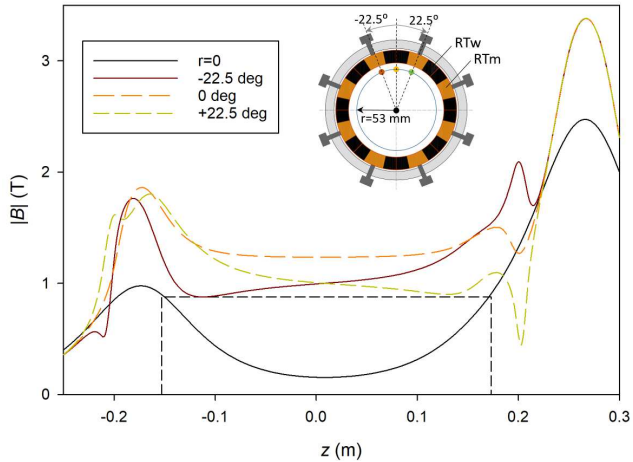}
    \caption{Magnetic field profiles along several vertical lines parallel to the IPT axis ($z$-axis) at the distance of 53 mm from the trap axis. The locations of these lines in the horizontal cross-section are shown in the insert. Vertical dashed lines mark locations of the top and bottom of the trapping cell. Plots are calculated for  100$\%$ of $I_t$ (0.8 T trap depth).}
    \label{z_profiles}
\end{figure}
Calculations of the magnetic field were performed at the stage of designing the IPT for various configurations of its components and their mutual arrangement in order to reach the specified parameters. We have not performed a detailed measurement of the field profile in the IPT using e.g. a Hall probe. This would require a complicated probe positioning system. The calculations based on the Biot-Savart law provide reliable results, and a simple check for a few points in the IPT was considered to be sufficient. These checks were performed as described above during the coil tests in liquid helium and their results agreed well with the calculations. In Fig.\ref{z_profiles} the axial magnetic field profiles are shown for the 80$\%$ of target currents, reached in the tests at 1.5 K. 

\begin{figure}[t]
    \includegraphics[width=\columnwidth]{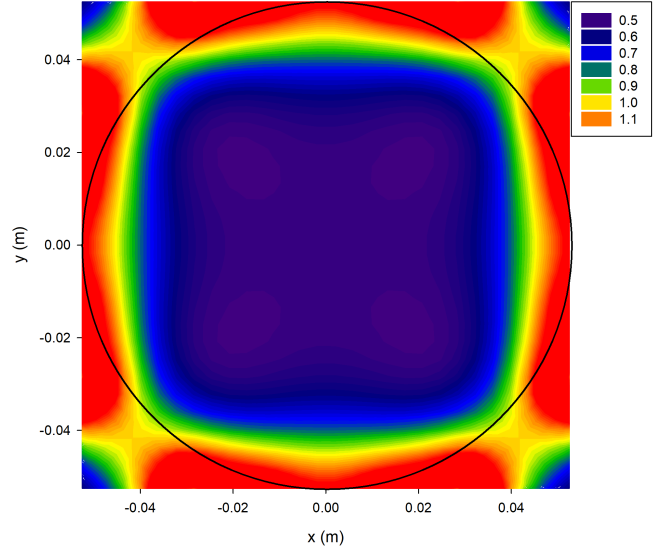}
    \caption{2D colour plot of magnetic field in the cross-section at $z=$-113 mm where the trap minimum (leakage spot) is located. Colour legend provides field in the range 0.5-1.1 T. The black circle corresponds to the trapping cell inner surface at $=53$ mm. Plots are calculated for the 100$\%$ of $I_t$ (0.8 T trap depth).}
    \label{surface_plot_leakage}
\end{figure}

Special attention was paid to find the trap leakage spots which define the minimum trap depth. As described below they are located above the LP and below the UP in the regions where the radial component of the pinch coils adds destructively to the field of the octupole. For the lower/upper pinches, these spots appear at the distance $z\approx-113/+131$ mm from the center of the IPT. A 2D colour plot for the crossection going through the lower pinch leaking location is presented in Fig. \ref{surface_plot_leakage}.
From the plots in Figs. \ref{z_profiles} and \ref{surface_plot_leakage} one can see that the magnetic field $B_{min}^{wall}\approx0.95$ T in the leaking points on the inner surface of the trapping cell ($r=53$ mm) was reached during a steady operation of the IPT at 4.2 and 1.5 K. Taking into account the $B_{min}\approx0.15$ T field in the trap center, this justifies that the trap depth of $\Delta B_{t}\approx$ 0.8 T was reached. The field at the inner bore of the IPT ($r=56$ mm) is $\approx$1.1 T corresponding to the trap depth of $\approx$ 0.95 T.

\section{Conclusions and possible improvements}
We attained the trap parameters specified above for experiments with atomic hydrogen. The volume and field depth of the trap are sufficient to confine $>10^{15}$ hydrogen atoms at 50 mK. This can be reached at low densities $\sim10^{12}$ cm$^{-3}$ when the decay of the sample caused by dipolar relaxation occurs on the time scale of $\sim$10 min. Our trap performed equally well in a bath of liquid helium at 4.2 K and inside the vacuum can of the dilution cryostat at 1.5 K. This indicates that the performance is not limited by the critical currents of the superconductors, which increase by $\approx30\%$ at lower operating temperature. The mechanical stability of the IPT assembly against magnetic forces is a major concern for large size traps, and it is clear that it is the main limitation also here. We found that the construction of the octupole magnet built with eight race-track coils pressed from outside by metal rings, as the staves of a wooden barrel, helps to improve the mechanical stability. This design also allows an easy disassembling of the IPT and replacing of any of its components including individual race-track coils.

The maximum trap depth is determined by the difference of the magnetic field minimum on the inner surface of the trapping cylinder and the field $B_{min}$ at the trap bottom. The field $B_{min}\approx0.15$ T reached in the tests of our IPT is fairly large and is not required for the experiments with trapped H. It is a consequence of the short distance between the pinch coils constrained by the overall IPT length. A simple solution to reduce $B_{min}$ is to place an extra trim coil in the center of the IPT between the pinches producing a field of opposite direction to that of the pinches. Cancellation of the 0.15 T field to nearly zero can be done with a relatively weak solenoid which will not change significantly the field in the leakage spots. This simple modification would deepen the trap by $\approx15\%$. 

The maximum ramping rates of the IPT currents are limited now by a poor thermal contact between the trap and the 1 K pot of the refrigerator. There are several ways to improve this thermal link. We found that the main bottleneck in the heat transfer is the small contact area between the copper rods serving as thermal links and the liquid helium inside the pot. The main function of the 1 K pot of a dilution refrigerator is to condense return $^3$He using a dedicated heat exchanger inside the pot. Normally, the bottom of 1 K pot is not designed to transfer large amounts of heat. This can be improved by installing copper rods from the IPT into the pot and increasing the contact area with liquid $^4$He using sintered heat exchangers. Another possibility is to transfer heat from the IPT using tubes filled with superfluid helium instead of the copper rods.

A continuous operation of the IPT at full current will become possible after simple improvements of the current leads. Our cryostat has no liquid nitrogen bath, the current leads go through its neck and are cooled by helium vapour. The natural boil-off of our cryostat of $\approx0.5$ l/h is rather low. With the full currents in the IPT leads the boil-off increases by $\sim50\%$, which is still not sufficient for cooling the leads. An obvious solution is to increase the cross-section of the leads between the room temperature and the HTS tapes. This will increase the flux of helium vapour in the neck and improve cooling of the leads at the expense of the higher helium boil rate. This would not be a problem in modern dry dilution refrigerators utilizing two or three cryocooleres with cooling powers up to 1.5 W at 3 K. In fact the cooling power of a single cryocooler is twice as large as the current lead heating in our tests described above. One should also point out that the cooling power of the cryocoolers is very large at their 40 K stage, to which the ends of the HTS tapes could be thermally pinned. 

The possibility of operating in vacuum is especially valuable for the integration of the IPT into the standard dry dilution cryostat where there is no $^4$He bath.  The large diameter of the vacuum space of a typical dry cryostat and the location of the plates where the 40 K and 4 K temperatures are reached, is much more convenient for installing the IPT allowing simpler and shorter thermal links. 

In conclusion, we developed a large Ioffe-Pritchard trap for desktop experiments with atomic hydrogen. The trap has the largest volume and about the same strength as earlier traps used for cold atoms or anti-atoms. The trap construction allows an easy replacement of any of its components, including the race-track coils of the octupole magnet. Our IPT operates in vacuum and can be easily integrated in modern dry dilution refrigerator cryostats operating with pulse tube cryocoolers.

\section{Acknowledgements} 
This work was supported by the Academy of Finland (Grant No. 317141), the Jenny and Antti Wihuri Foundation, International Emerging Action (IEA00552) "Quantum Reflection of Cold Hydrogen", CNRS. The authors would like to thank M. W. Reynolds (Ripplon Software Inc.) for providing BiotSavart Magnetic Field Calculator, Rapha\"{e}l Janiv (Teijin Frontier Europe GMBH) for help with Zylon filament. We also thank Simo Jaakkola for commenting the manuscript.

\bibliographystyle{apsrev}
\bibliography{IPT}
\end{document}